\documentclass[aps,prl,twocolumn,amsmath,amssymb,superscriptaddress]{revtex4}%
\usepackage{graphicx}
\usepackage{dcolumn}
\usepackage{bm}
\usepackage{color}
\usepackage{amsmath}
\usepackage{epstopdf}
\usepackage{fancybox}
\usepackage{amsfonts}
\usepackage{amssymb}%
\usepackage[toc,page]{appendix}
\renewcommand{\cite}[1]{{[}\onlinecite{#1}{]}}

\newcommand{\be}{\begin{equation}}
\newcommand{\e}{\end{equation}}
\newcommand{\beml}{\begin{subequations}}
\newcommand{\eml}{\end{subequations}}
\newcommand{\beq}{\begin{eqnarray}}
\newcommand{\eq}{\end{eqnarray}}
\newcommand{\ba}{\begin{array}}
\newcommand{\ea}{\end{array}}
\newcommand{\bpm}{\begin{pmatrix}}
\newcommand{\epm}{\end{pmatrix}}
\newcommand{\bc}{\begin{cases}}
\newcommand{\ec}{\end{cases}}

\begin{document}

\title{Ultrafast generation and dynamics of isolated skyrmions in antiferromagnetic insulators}

\author{Rohollah Khoshlahni}
\affiliation{Institute for Advanced Studies in Basic Science (IASBS), Zanjan, Iran}
\author{Alireza Qaiumzadeh}
\thanks{Corresponding author: alireza.qaiumzadeh@ntnu.no}
\affiliation{Center for Quantum Spintronics, Department of Physics, Norwegian University of Science and Technology, NO-7491 Trondheim, Norway}
\author{Anders Bergman}
\affiliation{Division of Materials Theory, Department of Physics and Astronomy, Uppsala University, Box 516, 75120 Uppsala, Sweden}
\author{Arne Brataas}
\affiliation{Center for Quantum Spintronics, Department of Physics, Norwegian University of Science and Technology, NO-7491 Trondheim, Norway}

\begin{abstract}
Based on atomistic spin dynamics simulations, we report the ultrafast generation of single antiferromagnetic skyrmions in a confined geometry. This process is achieved through an effective magnetic field induced by the athermal inverse Faraday effect from a short laser pulse. The resulting field can nucleate an isolated skyrmion as a topologically protected metastable state in a collinear antiferromagnet with small Dzyaloshinskii-Moriya interaction. The radius of a single skyrmion is shown to increase by applying a uniform dc magnetic field and at increasing temperature. To investigate possible AFM spin-caloritronics phenomena, we investigate the skyrmion dynamics under an applied temperature gradient both analytically and numerically. The antiferromagnetic skyrmions move longitudinally toward the hotter region, but to in contrast, small skyrmions in the very low damping regime move toward the colder side, irrespective of the staggered topological charge number, with a speed that is much faster than that of their ferromagnetic counterparts.
\end{abstract}

\date{\today}
\maketitle
\section{Introduction}
Antiferromagnetic (AFM) spintronics is an emerging and fast-growing subfield in spintronics that promises faster, smaller and more energy efficient state-of-the-art memory devices and data processors \cite{RevModPhys.90.015005, Rev-Nature, gomonay2018antiferromagnetic,jungwirth2016antiferromagnetic,gomonay2014spintronics,macdonald2011antiferromagnetic,marti2015prospect}.
The dynamics of AFM systems are more complicated than that of their ferromagnetic (FM) counterparts and exhibit richer physics. Despite being discovered as early as the 1930s \cite{neel1936proprietes,bitter1938generalization}, the absence of a net magnetization and the associated insensitivity to magnetic fields \cite{jungfleisch2018perspectives} have hitherto limited the use of antiferromagnets. The only use for antiferromagnets is in passive exchange-bias structures.
With recent advances in experimental techniques, as well as novel theoretical proposals, the door to the AFM spintronics era has opened a little further \cite{doi:10.1002/pssr.201700022}. Important observations and predictions are unprecedented long-range spin transport in AFM insulators \cite{Alireza-Nature}, detection and manipulation of the N\'{e}el order \cite{Electricalswitching,PhysRevLett.108.247207}, engineering of AFM domain walls (DWs) \cite{logan2012antiferromagnetic}, and AFM-DW motion \cite{DWmotion,PhysRevLett.117.107201,PhysRevLett.110.127208}.

The Dzyaloshinskii-Moriya interaction (DMI) is an antisymmetric exchange interaction of a relativistic origin that breaks the chiral symmetry in magnetic systems \cite{moriya1960anisotropic,dzyaloshinsky1958dzyaloshinsky}. Initially, the DMI was identified as the mechanism responsible for the weak magnetism observed in a few AFM systems, namely, the so-called \textit{weak} FM systems. In general, within the continuum limit, the DMI decomposes into two parts in AFM systems: one being homogeneous, and the other being inhomogeneous. Whether these parts are finite depends on the underlying crystallographic symmetry of the AFM system. The homogeneous DMI is responsible for weak ferromagnetism \cite{moriya1960anisotropic} while the finite inhomogeneous part breaks the chiral symmetry and stabilizes exotic spin textures with well-defined chirality, such as chiral DWs and helimagnets \cite{bogdanov2002magnetic,Alireza-DMI}.

Skyrmions, which are nanoscale swirling magnetic textures, are topologically invariant chiral solitons. The inhomogeneous DMI can stabilize skyrmions in magnetic systems with broken inversion symmetry. Although these solitons were predicted quite a long time ago, the experimental observation and creation of skyrmions occurred only recently in FM systems, either as skyrmion lattices or as single skyrmions \cite{Polyakov,bogdanov1989thermodynamically,bogdanov1994thermodynamically,bogdanov1994properties,bogdanov1999stability,tchoe2012skyrmion,flovik2017generation,liu2015skyrmion}. Single skyrmions can be utilized in encoding, transmitting and processing information in spintronic devices \cite{fert2013skyrmions,zhang2015magnetic,zhou2014reversible}. Thus far, skyrmions have been observed only in FM and long-wavelength spin spiral systems.  Recently, there have been predictions that it is possible to stabilize these topological solitons even in AFM systems as either skyrmion lattices or isolated skyrmions \cite{zhang2016antiferromagnetic,raivcevic2011skyrmions,gobel2017antiferromagnetic,RembertSkyrmion,AFMSkyrmion1,AFMSkyrmion2,AFMSkyrmion3,jin2016dynamics,velkov2016phenomenology,barker2016static,bessarab2017stability,AFMIan}.

To date, there have been only a few proposals for the generation and control of isolated skyrmions in AFM systems. Spin-transfer torques induced by spin (polarized) currents can create skyrmions \cite{AFMSkyrmion2,zhang2016antiferromagnetic,AFMracetrack}, and spin (polarized) currents can be applied to move them \cite{AFMSkyrmion2,velkov2016phenomenology,barker2016static,jin2016dynamics,AFMracetrack}. These proposals for the creation and control of AFM skyrmions have some limitations and drawbacks. For example, some of them only apply to metallic AFM systems. Furthermore, all of the proposed methods depend on the use of heterostructured materials, and more importantly, the incubation time for the generation of a single AFM skyrmion is also long,  a few nanoseconds \cite{zhang2016antiferromagnetic}.

In this paper, we propose a method for the ultrafast generation of
 single AFM skyrmions in a confined geometry employing an effective magnetic field induced by the optical inverse Faraday effect (IFE) \cite{Rasing}. We also study the AFM skyrmion motion induced by the magnonic Seebeck effect numerically in an atomistic spin dynamic simulation and analytically by using a collective coordinate approach. Thus, our method can be used to generate and move isolated skyrmions in single crystals of AFM insulators. We organize the remainder of this paper as follows. In Sec. II, we introduce our AFM system and the equations of motion for AFM spins. In Sec. III, we present our results for the rapid generation of single AFM skyrmions. We discuss the dynamics of isolated skyrmions in the presence of thermal magnons in Sec. IV. Finally, we conclude the paper in Sec. V and discuss the outlook on future work.

\begin{figure}[t]
	\hspace{-0.9 cm}
	\begin{minipage}[b]{0.5\linewidth}
		\centering
		\includegraphics[width=1.2\linewidth]{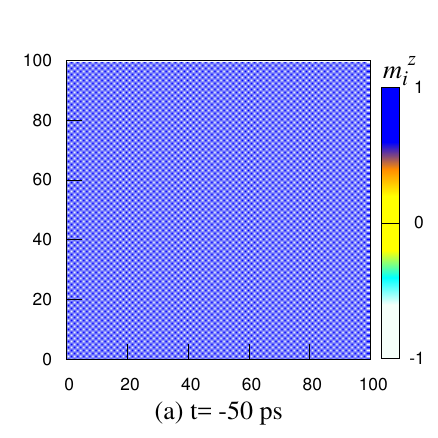}
	\end{minipage}
	\hspace{0.5cm}
	\begin{minipage}[b]{0.5\linewidth}
		\centering
		\includegraphics[width=1.2\linewidth]{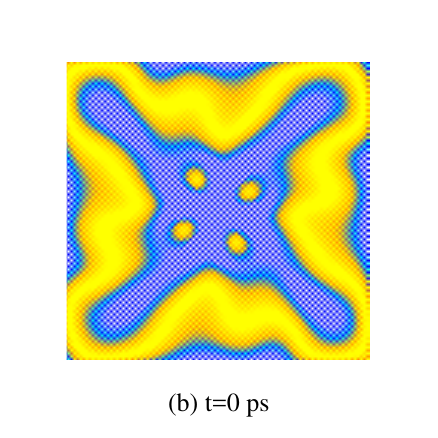}
		\end{minipage}
	\hfill
	\begin{minipage}[b]{0.52\linewidth}		
		\centering
		\hspace{-1.3 cm}
		\includegraphics[width=1.2\linewidth]{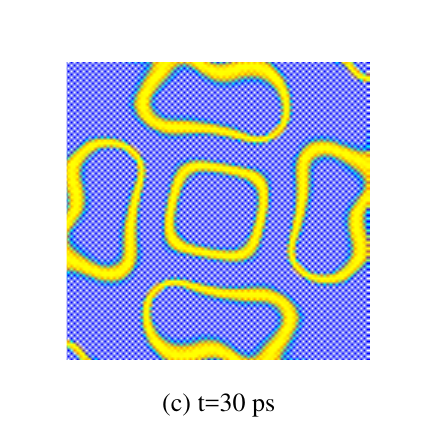}
		\end{minipage}
	\hspace{-0.5cm}
	\begin{minipage}[b]{0.51\linewidth}
		\centering
		\includegraphics[width=1.2\linewidth]{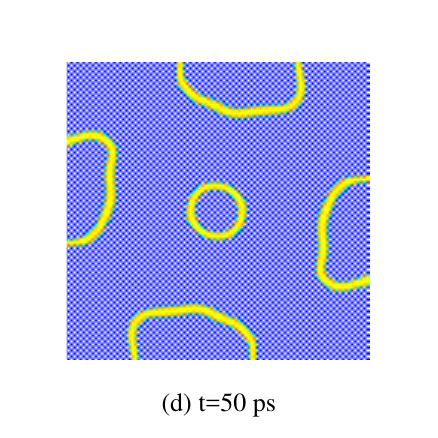}
			\end{minipage}
		\hfill
		\begin{minipage}[b]{0.46\linewidth}		
			\centering
			\hspace{-0.88 cm}
			\includegraphics[width=1.2\linewidth]{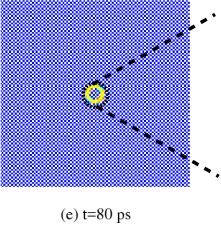}
		\end{minipage}
		\begin{minipage}[b]{0.4\linewidth}
			\centering
			\includegraphics[width=1.17\linewidth]{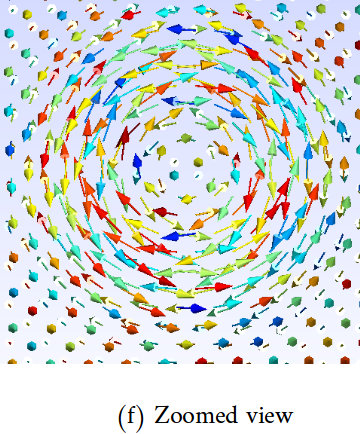}
		\end{minipage}
	\caption{(Color online) Snapshots of the time evolution of the spin configuration induced by a single 30 ps Gaussian magnetic field pulse normal to a square monolayer. (a) The initial state is an AFM ground state. (b) The maximum peak of a Gaussian magnetic field pulse arrives at t = 0, and a domain with $M_z=0$ starts to form. (c) Evolution of a domain wall to create a preliminary design of AFM skyrmion. (d) Domains shrink, and some reach the boundary and disappear. The remaining domains form a circle in the center. (e) Ultimately, one chiral skyrmion is stabilized in the center of the monolayer. (f) A magnified view of a chiral AFM skyrmion.}
\label{snapshots}
\end{figure}

\section{AFM Hamiltonian and dynamics}
We consider a discrete bipartite two-dimensional (2D) AFM insulator with the following effective thermodynamic free energy:
\begin{align} \label{Hamiltonian}
\mathcal{F}=&-\sum_{\langle i,j\rangle}J_{ij}\bm{m}_{i}\cdot\bm{m}_{j}-\sum_{\langle i,j\rangle}\bm{D}_{ij}\cdot\bm{m}_{i}\times \bm{m}_{j}\nonumber\\
&+K\sum_{i}(\bm{m}_{i}\cdot\hat{z})^2-\mu_{s}\sum_{i}\bm{h}(t)\cdot\bm{m}_{i},
\end{align}
where $\bm{m}_i$ is the unit vector of the spin magnetic moment at site $i$. On the right-hand side of Eq.\ ({\ref{Hamiltonian}}), the first term is the Heisenberg exchange interaction, with $J_{ij}<0$ representing the nearest-neighbor AFM exchange energy; the second term is DMI, with the DMI vector $\bm{D}_{ij}$; the third term is the single ion anisotropy in the $z$ direction, with $K<0$ being the uniaxial anisotropy energy; and the last term is the Zeeman interaction between the external time-dependent magnetic field $\bm{h}$ and the localized spins, with $\mu_s$ being the sublattice saturation magnetization.

The Heisenberg exchange interaction forces adjacent spins to become antiparallel, whereas the DMI encourages perpendicular configurations of neighboring spin moments. The competition between these two energy scales leads to various exotic spin textures in the ground state or metastable states \cite{fert2013skyrmions,woo2016observation}. When the DMI strength is larger than a critical value, $D>D_c=4\sqrt{J K}$, \cite{moon2013spin} the ground state differs from a collinear AFM state. In simple square lattices, there are two types of DMIs based on the DM vector alignment \cite{chaurasiya2016direct}. We denote DMI as bulk (interfacial) DMI when the DM vector is parallel (perpendicular) to the bond direction. The bulk DMI is responsible for textures with Bloch-like structures in noncentrosymmetric crystals while the interfacial DMI leads to N\'{e}el-like structures at either the interface of heavy metals and AFM bilayers or AFM systems with broken inversion symmetry \cite{Alireza-DMI}. In this paper, we present the results for the bulk DMI. An extension of our results to the interfacial DMI is possible. In the free energy ({\ref{Hamiltonian}}), we disregard the long-range dipolar interactions since they are negligible in thin film of AFM systems. We also assume that the temperature is much less than the N\'{e}el temperature. In this limit, we treat spins as 3D vectors with a fixed length, $|\bm{m}_{i}|=1$.

The dynamics of atomic moments in an AFM system are described by the stochastic Landau-Lifshitz-Gilbert (sLLG) equation \cite{garcia1998langevin,skubic2008method},
\begin{equation}\label{LLG}
\frac{d\boldsymbol{m}_{i}}{dt}=-\tilde{\gamma}\boldsymbol{m}_{i}\times\left[(\bm{H}_i+\bm{H}_i^{\mathrm{th}})+\alpha_\mathrm{G} \boldsymbol{m}_{i}\times(\bm{H}_i+\bm{H}_i^{\mathrm{th}})\right],
\end{equation}
where $\tilde{\gamma}=\gamma/(1+{\alpha}^2)$ is the renormalized gyromagnetic ratio, $\alpha_\mathrm{G}$ is the effective Gilbert damping parameter, $\bm{H}_i=-\partial\mathcal{F}/(\mu_s\partial{\boldsymbol{m}_{i}})$ is the effective magnetic field on site $i$, and $\bm{H}_i^{\mathrm{th}}$ is the stochastic magnetic field arising from the thermal fluctuations.
The stochastic magnetic field describes how temperature effects enter the theory of atomistic spin dynamics in a Langevin dynamics approach.
Using the fluctuation-dissipation theorem, the thermal stochastic fields can be described by the following correlations that are local in both space and time:
\begin{align}
\langle H^{\mathrm{th}}_{i,\alpha}(t) H^{\mathrm{th}}_{j,\beta} (t')\rangle&=2 \xi_H \delta_{ij}\delta_{\alpha\beta}\delta(t-t'),\label{Correlators1}\\
\langle H^{\mathrm{th}}_{i,\alpha}(t) \rangle&=0,
\end{align}
where $\xi_H=\alpha_\mathrm{G} k_\mathrm{B} T/(\gamma \mu_s)$ is the noise power \cite{kubo1970brownian}. Throughout this paper, we use Latin letters for site numbers and Greek letters for the spatial components of a vector. In Eq. (\ref{Correlators1}), the quantum effects that appear at lower temperatures have been ignored.
Performing atomistic spin dynamic simulations, we solve the sLLG equation, Eq. (\ref{LLG}), using the Uppsala Atomistic Spin Dynamics (UppASD) code \cite{skubic2008method,UppASD}.

\section{Ultrafast generation of isolated AFM skyrmions}

Skyrmions appear either in the skyrmion crystalline phase in a stable state or as isolated skyrmions in a metastable state. Isolated skyrmions are central for data storage and processing. Hence, controlling single skyrmions is essential for practical applications. In this section, we propose an ultrafast method to create single skyrmions in confined geometries. Creating a single skyrmion in a metastable state requires transforming the system from the ground state, i.e., the collinear state, into a new local minimum containing a skyrmion state. Here, we show that applying an intense and short magnetic field pulse can create single skyrmions in AFM insulators via magnon instability processes \cite{flovik2017generation}.

\begin{figure}[t]
\centering
\includegraphics[width=0.5\textwidth]{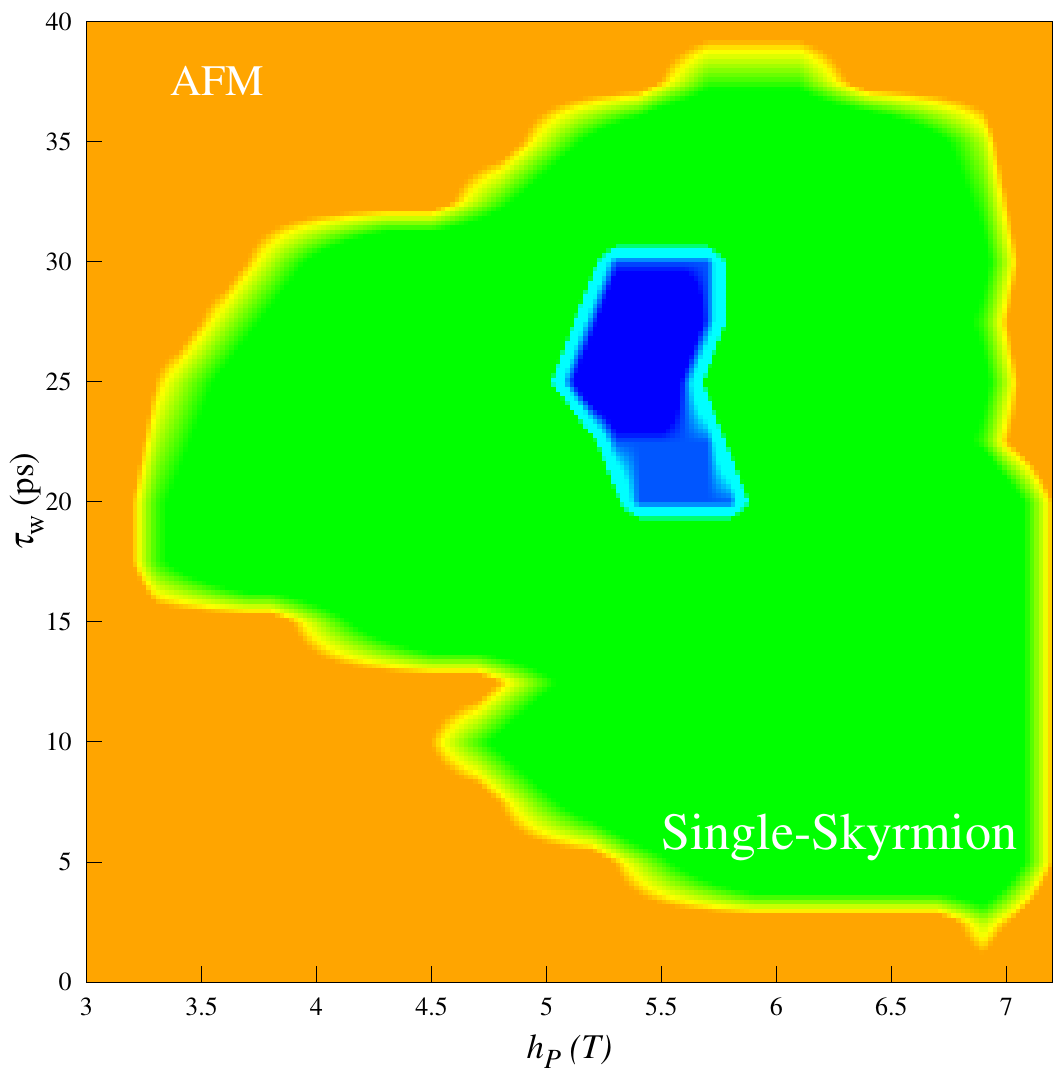}
\caption{(Color online) Phase diagram for the skyrmion nucleation by applying a magnetic field pulse on a square with a size of $100d \times 100d$. The sand color shows the AFM ground state. The green region represents the isolated skyrmion metastable state, which survives even after turning the uniform and dc magnetic field off. The blue color shows the isolated skyrmion metastable state, which exists only in the presence of an external uniform magnetic field.}
\label{phase-space}
\end{figure}

The recent discovery of ultrafast and nonthermal magnetization dynamics triggered by intense and polarized laser pulses has attracted attention and promises a new route toward ultrafast opto-magnetism \cite{kimel2005ultrafast,kimel2007nonthermal,koopmans2003laser}.
Although the underlying theory behind this effect is still unclear, phenomenologically, the effect of a polarized laser on magnetic systems is to produce an effective magnetic field induced by the IFE $\bm{h}\propto \bm{E}(t)\times \bm{E}^*(t)$, where $\bm{E}$ is the electric field of a laser pulse \cite{Rasing}. The amplitude of the magnetic field is proportional to the light intensity, its sign depends on the helicity of the pulse, and its direction is along the light propagation.

There are recent reports of ultrafast optical nucleation of single skyrmions and skyrmion lattices in ferrimagnetic and ferromagnetic materials using laser pulses, but the microscopic origin is attributed to laser-induced transient heating \cite{finazzi2013laser,BubbleLattices,PhysRevLett.120.117201}. The possibility of the creation of skyrmions using optical vortex beams, electromagnetic waves carrying intrinsic orbital angular momentum, has theoretically been investigated recently \cite{OAM1,OAM2}.
In this paper, we are interested in the nonthermal effects of circularly polarized laser pulses caused by the IFE \cite{Rasing} in a confined AFM system with an initial collinear state, i.e., $D<D_c$. We model the light-induced effective magnetic field or IFE by a time-dependent Gaussian magnetic field pulse in the sLLG equation, $\bm{h}(t)=h_p\mathrm{exp}(-t^2/2\tau_{w}^2) \hat{z}$, where $h_p$ is the pulse amplitude and $\tau_w$ is the pulse width. The amplitude of this effective magnetic field can be a few tesla, and its effective duration is subnanosecond \cite{Blugel,Koopmans}.

We consider a confined square lattice of $100d\times100d$ spins, where $d=3\AA$ is the lattice constant. The Heisenberg exchange interaction is isotropic $J_{ij}=J$, as is the DMI, $|\bm{D}_{ij}|=D$. We choose typical material parameters in our atomistic spin dynamics simulations: the AFM exchange energy $J=-0.5$ meV/atom, $K=0.1$ J, $D=0.15$ J and $\alpha_\mathrm{G}=0.009$. Using UppASD, we find that the ground state of the system is a collinear AFM state with tilted spins at the boundaries due to the competition between DMI and exchange energy; see Fig. \ref{snapshots}-a and the Supplementary Material \cite{suppl}.

Next, we apply a magnetic field pulse with $h_p=9$ T and $\tau_{w}=30$ ps normal to the sample. Magnons with different wavevectors are excited at the boundaries and propagate inside the system. Figure \ref{snapshots}-b shows that when the magnetic field pulse reaches its maximum, several skyrmion nuclei form in the middle of the system. After recombination and repulsion of the nuclei, a single skyrmion survives at the center of the sample; see Fig. \ref{snapshots}-e. Figure \ref{snapshots}-f shows that this AFM skyrmion, as expected, is of a Bloch-type since the DMI is bulk-type and isotropic in our square lattice structure. We have also checked the effect of the next-nearest-neighbor exchange interaction and observed a similar skyrmion nucleation process, as depicted in Fig. \ref{snapshots}, but at a slightly smaller applied magnetic field with the same pulse duration.

The application of a dc magnetic field normal to the sample can reduce the critical amplitude of the magnetic field pulse. The physical mechanism behind this reduction is that the barrier between the global minimum, the AFM collinear state, and the local minimum, the isolated skyrmion state, dramatically decreases in AFM systems near the so-called spin-flop phase. To find the phase diagram for isolated skyrmion nucleation, i.e., $\tau_w$ vs. $h_p$, we turn on a dc magnetic field of $h_0=5$ T, which is smaller than the spin-flop field of the system $\sim 7$ T, before applying magnetic field pulses of different amplitudes and durations. After turning off the dc magnetic field, at the end of the skyrmion incubation process, we check whether the final skyrmions are stable; see Fig. \ref{phase-space}. This phase diagram shows that it is possible to reduce the applied magnetic field by a few teslas. Within the phase diagram, there is a region, shown in blue, in which isolated skyrmions are stable only in the presence of a dc magnetic field and disappear by switching off the magnetic field.
Note that both thresholds of pulse duration and amplitude for skyrmion nucleation are very material dependent.

\begin{figure}[t]
\centering
		\hspace{-.5 cm}
	\includegraphics[width=.455\textwidth]{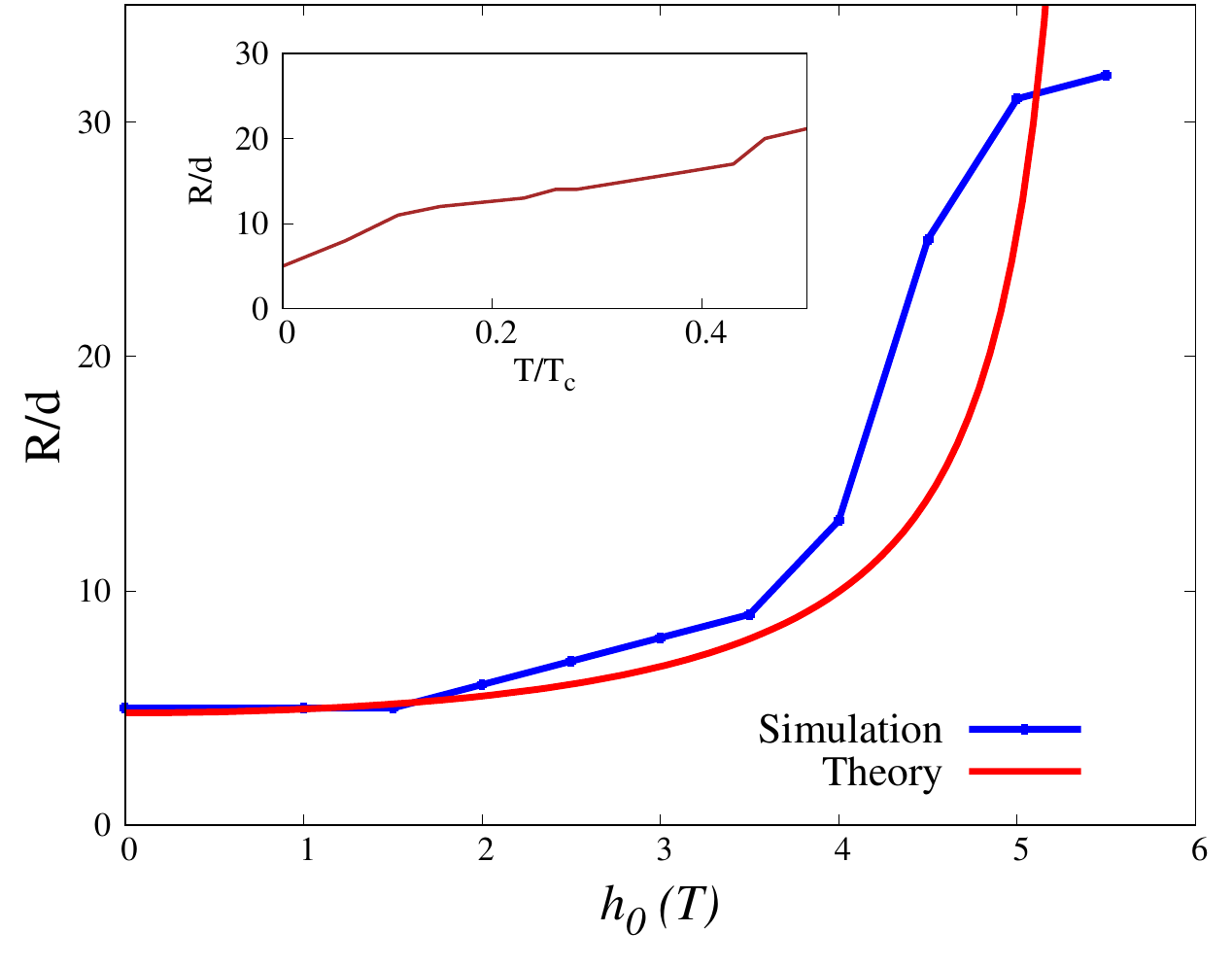}
	    \caption{(Color online) Skyrmion radius versus magnetic field at zero temperature. The red solid curve represents the analytical prediction, and the blue solid curve results from the atomistic simulations. The inset shows that the AFM skyrmion radius increases with temperature.}
\label{R-B}
\end{figure}

The Zeeman energy arising from the coupling of an external magnetic field with local magnetic moments appears to be an effective hard-axis anisotropy term in the free energy of AFM systems expressed as a function of the N\'{e}el vector \cite{Alireza-SSF}. It is possible to demonstrate that the radius of AFM skyrmions in the regime $D<D_c$  always increases with an applied dc magnetic field, irrespective of the magnetic field sign, $R/d = -\pi D/\left(K + \mu_s^2 B^2 /(16 |J|)\right)$ \cite{bessarab2017stability,Bogdanov}. This feature differs from FM systems, where the sign of the magnetic field controls the skyrmion size $R/d=\pi D/(K + 8\mu_s B /\pi^2)$ \cite{Bogdanov,SkyrmionSize}.
Figure \ref{R-B} presents the variation in the AFM skyrmion radius as a function of an applied perpendicular dc magnetic field. The AFM skyrmion size increases with magnetic field irrespective of the direction of the magnetic field, which is different from FM skyrmions \cite{tomasello2018origin}. Figure \ref{R-B} shows good agreement between the results of atomistic simulations and the theory \cite{bessarab2017stability}. In the inset of Fig. \ref{R-B}, we show that the radius of AFM skyrmions increases with temperature, as has already been predicted theoretically \cite{barker2016static}.

\section{AFM skyrmion motion induced by magnonic Seebeck effect}
The application of skyrmions as data bits in racetrack memories requires their motion to be deterministic. In AFM insulators, recent theories suggest that either coherent \cite{Alireza-SW1, Alireza-SW2} or incoherent (thermal) magnons \cite{PhysRevLett.117.107201} drive domain wall motion. Traveling incoherent magnons can be excited by applying a thermal gradient across the AFM system. Magnons in AFM systems, contrary to their FM counterparts, possess either left- or right-handed circular polarizations with opposite spin angular momenta. At finite temperatures, both species of magnon polarizations are excited with an equal population such that thermal magnons carry no net spin angular momentum.

In this section, we explore the dynamics of single AFM skyrmions under a thermal gradient.
First, we derive a theory for the motion of AFM solitons in the presence of a thermal gradient at the continuum level, and then we present our atomistic simulations.

\subsection{Stochastic LLG equation for N\'{e}el vector dynamics}

We consider a two-sublattice AFM insulator in the continuum limit, i.e., $d \rightarrow 0$. At low temperature, the magnetic moment in sublattices are $\bm{m}_A$ and $\bm{m}_B$, where $|\bm{m}_A|=|\bm{m}_B|=1$. For analytic calculations, it is more convenient to introduce two new variables: a total magnetization field inside the unit cell $\bm{m}=\bm{m}_A+\bm{m}_B$ and a staggered order parameter $\bm{n}=(\bm{m}_A-\bm{m}_B)/|\bm{m}_A-\bm{m}_B|$, where $\bm{m}\cdot\bm{n}=0$ and $\bm{n}=1$.
The total AFM Lagrangian, $\mathcal{L}=\mathcal{L}_{\mathrm{kin}}-\mathcal{F}$, is the difference between the kinetic energy $\mathcal{L}_{\mathrm{kin}}$ and the thermodynamic free energy $\mathcal{F}$,
\begin{align}
\label{AFM-Lag}
&\mathcal{L}_{\mathrm{kin}}=\int d^2\bm{r}\frac{1}{2a} \dot{\bm{n}}^2,\\
&\mathcal{F}=\int d^2\bm{r}\left(\frac{A}{2}(\nabla \bm{n})^2+\frac{D}{d} \bm{n}\cdot(\nabla\times \bm{n})\right),\label{AFM-Lag1}
\end{align}
where $a$ and $A$ are the homogeneous and inhomogeneous exchange stiffnesses, respectively, and $D$ is the inhomogeneous DM coefficient. It is straightforward to show how the energy parameters in the continuum model, Eqs. (\ref{AFM-Lag}) and (\ref{AFM-Lag1}), are related to the energy parameters in the discrete model, Eq. (\ref{Hamiltonian}) (e.g., see Ref. \cite{Erlend}).  Minimizing the total Lagrangian in the presence of  dissipation, using a Rayleigh dissipation function $\mathcal{R}=(\mu_s/\gamma)\alpha_\mathrm{G} \int d^2\bm{r}\dot{\bm{n}}^2/2$, we obtain
\begin{align}
\label{LLG}
\bm{n}\times(\ddot{\bm{n}}-a \bm{f}_{\bm{n}}+\frac{\mu_s}{\gamma}a\alpha_\mathrm{G} \dot{\bm{n}})=0,
\end{align}
where $\bm{f}_{\bm{n}}=-\delta \mathcal{F}/\delta \bm{n}$ is the effective staggered field.

The inclusion of finite temperature effects is via the Langevin dynamics by adding a stochastic Gaussian-shaped field $\bm{f}^{\mathrm{th}}$ to the effective staggered field. Then, the sLLG equation becomes,
\begin{align}
\label{sLLG}
\bm{n}\times\left(\ddot{\bm{n}}-a (\bm{f}_{\bm{n}}+\bm{f}^{\mathrm{th}})+\frac{\mu_s}{\gamma}a\alpha_\mathrm{G} \dot{\bm{n}}\right)=0,
\end{align}
The dissipation-fluctuation theorem relates the Langevin field to the damping constant,
\begin{align}
\label{diss-Fluc}
&\langle f_\alpha^{\mathrm{th}}(\bm{r},t) f_\beta^{\mathrm{th}} (\bm{r}',t') \rangle=2\xi \delta_{\alpha\beta} \delta(\bm{r}-\bm{r}') \delta(t-t'),\\
&\langle \bm{f}^{\mathrm{th}}(\bm{r},t) \rangle=0,
\end{align}
where $\xi=\alpha_\mathrm{G} k_B T(x)$ is the correlation amplitude.

We can introduce two length scales: one is the helix wavelength $\Delta \equiv d(A/D)$, and the other one is the thermal-magnon wavelength $\lambda_T\propto d \sqrt{A/(k_B T)}$. Throughout our calculations, we assume $\Delta\gg\lambda_T$, which is valid for thermal magnons.

\subsection{Effective sLLG equation of AFM soliton}

To derive an effective description of the skyrmion dynamics, we introduce fast spin fluctuations $\delta\bm{n}$ generated by thermal fluctuations around a slowly varying magnetic texture $\bm{n}^{(0)}$,
\begin{align}
\label{magnon}
\bm{n}=\sqrt{1-\delta \bm{n}^2}\bm{n}^{(0)}+\delta \bm{n},
\end{align}
where $\delta \bm{n}\cdot \bm{n}^{(0)}=0$.

Substituting Eq. (\ref{magnon}) into the sLLG equation (\ref{sLLG}) and integrating over the fast oscillating component, we find the effective stochastic equation of the motion,
\begin{align}
\label{effective-sLLG}
\bm{n}^{(0)}\times\left(\ddot{\bm{n}}^{(0)}-a \bm{f}^{\mathrm{th}}+\frac{\mu_s}{\gamma}a\alpha_\mathrm{G} \dot{\bm{n}}^{(0)}\right)+\bm{\tau}^{\mathrm{magn}}=0,
\end{align}
where the thermomagnonic torques are given by,
\begin{align}
\label{therm-torque}
\bm{\tau}^{\mathrm{magn}}&=-a A \left( \langle\delta\bm{n} \times \partial^2_i \delta \bm{n}\rangle - \partial_i \langle\delta \bm{n}^2\rangle \bm{n}^{(0)}\times \partial_i\bm{n}^{(0)} \right)\nonumber\\
&=-a \hbar \bm{J}^n\cdot\nabla\bm{n}^{(0)} + a A (\partial_i \rho) \bm{n}^{(0)}\times \partial_i\bm{n}^{(0)}.
\end{align}
where the AFM magnon current is $J_i^n=(A/\hbar) \bm{n}^{(0)}\cdot\langle\delta\bm{n} \times \partial_i\delta\bm{n}\rangle$, and the AFM magnon number density is $\rho=\langle\delta \bm{n}^2\rangle/2$.
The adiabatic thermomagnonic torque, Eq. (\ref{therm-torque}), in AFM systems has two contributions with opposite signs. The first term is a reactive torque, and the second one is a dissipative torque \cite{thermomagnonic,Thermophoresis,PhysRevB.97.054308}.

\subsection{Stochastic Thiele's equation}
To find a stochastic equation for the dynamics of AFM solitons, we follow Thiele's approach \cite{PhysRevLett.30.230}. We use collective coordinates for describing the position of the skyrmion center $\bm {u}(t)$ as $\bm{n}^{(0)}(\bm{r},t)=\bm{n}^{(0)}(\bm{r}-\bm {u}(t),t)$. Multiplying both sides of the effective sLLG equation, Eq. (\ref{effective-sLLG}), by $\bm{n}^{(0)}\cdot \partial_\alpha \bm{n}^{(0)} \times $, we obtain,
\begin{align}
\label{sThiele0}
&-\ddot{u}_\beta \partial_\beta \bm{n}^{(0)}\cdot \partial_\alpha \bm{n}^{(0)} + \dot{u}_\beta \dot{u}_\gamma (\partial_\beta \partial_\gamma \bm{n}^{(0)})\cdot \partial_\alpha \bm{n}^{(0)}\nonumber\\&
-a\partial_\alpha \bm{n}^{(0)}\cdot \bm{f}^{\mathrm{th}}-\mu_s \gamma^{-1} a \alpha_\mathrm{G} \dot{u}_\beta \partial_\alpha \bm{n}^{(0)}\cdot \partial_\alpha \bm{n}^{(0)}\nonumber\\&
-a \hbar {J}_\beta^n \bm{n}^{(0)}\cdot\partial_\alpha \bm{n}^{(0)}\times \partial_\beta \bm{n}^{(0)} \nonumber\\&+ a A (\partial_\beta \rho) \partial_\beta \bm{n}^{(0)}\cdot \partial_\alpha\bm{n}^{(0)}=0,
\end{align}
where we have used $\dot{\bm{n}}=-\dot{u}_\beta \partial_\beta \bm{n}$ and $\ddot{\bm{n}}=-\ddot{u}_\beta \partial_\beta \bm{n}+\dot{u}_\beta \dot{u}_\gamma \partial_\beta \partial_\gamma\bm{n}$.

After integrating over the spatial coordinates, we finally find the stochastic Thiele's equation for AFM skyrmions,
\begin{align}
\label{sThiele1}
M^{\alpha\beta}(\ddot{u}_\beta +\alpha_\mathrm{G} a \mu_s \gamma^{-1} \dot{u}_\beta)+F_\alpha^{\mathrm{th}}+F_\alpha^r+F_\alpha^d=0.
\end{align}
This equation is similar to Newton's equation of motion for the massive particles in a viscous medium, which is totally different from the massless dynamics of FM skyrmions \cite{PhysRevLett.30.230,kong2013dynamics,Mathias,velkov2016phenomenology}.

In Eq. (\ref{sThiele1}), the thermal, reactive and dissipative forces are respectively defined as,
\begin{align}
\label{forces}
&F_\alpha^{\mathrm{th}}=\frac{1}{\Delta^2}\int d^2\bm{r} \partial_\alpha \bm{n}^{(0)}\cdot \bm{f}^{\mathrm{th}},\\
&F_\alpha^\mathrm{r}=\frac{4\pi \hbar Q^n}{\Delta^2} \varepsilon^{\alpha\beta} {J}_\beta^n,\\
&F_\alpha^\mathrm{d}=-\frac{c^2}{\Delta^2} M^{\alpha\beta}\partial_\beta \rho,
\end{align}
where $Q^n=(1/4\pi)\int d^2 \bm{r}  \bm{n}^{(0)}\cdot(\partial_x\bm{n}^{(0)}\times\partial_y\bm{n}^{(0)})$ is the topological skyrmion number for the staggered field, $M^{\alpha\beta}=(a \Delta^2)^{-1}\int d^2 \bm{r} \partial_\alpha\bm{n}^{(0)}\cdot \partial_\beta\bm{n}^{(0)}$ is the symmetric AFM mass tensor, $\varepsilon^{\alpha\beta}$ is the 2D Levi-Civita symbol, and $c=\sqrt{a A}$ is the effective AFM magnon velocity in an isotropic medium. In perfectly circular skyrmions, $M^{\alpha\beta}=M \delta_{\beta\alpha}$. The thermal force satisfies the following relations:
\begin{align}
\label{forces}
&\langle F_\alpha^{\mathrm{th}}(\bm{u},t) F_\beta^{\mathrm{th}}(\bm{u}',t')\rangle=2\tilde{\xi}\delta_{\alpha\beta}\delta(\bm{u}-\bm{u}')\delta(t-t'),\\
&\langle F_\alpha^{\mathrm{th}}(\bm{u},t)\rangle=0,
\end{align}
where $\tilde{\xi}=(a M/\Delta^2)\xi$.

Here we should emphasize that in AFM systems, we can define another topological number for the magnetization field in each sublattice or {\it{magnetic}} topological charge $Q^m_{1(2)}=(1/4\pi)\int d^2 \bm{r}  \bm{m}_{1(2)}\cdot(\partial_x\bm{m}_{1(2)}\times\partial_y\bm{m}_{1(2)})$. Although the {\it{staggered}} topological charge, $Q^n$, is finite for AFM skyrmions, the total topological number related to the magnetization field vanishes $Q^m_{1}+Q^m_{2}=0$.

We are interested in the steady-state limit of Eq. (\ref{sThiele1}),
\begin{align}
\label{sThiele2}
\dot{u}_\alpha=-\frac{\gamma}{M \alpha_\mathrm{G} a \mu_s}(F_\alpha^{\mathrm{th}}+F_\alpha^\mathrm{r}+F_\alpha^\mathrm{d}).
\end{align}
The AFM soliton velocity is inversely proportional to the Gilbert damping coefficient. Consequently, we expect a faster motion compared to FM solitons since the damping coefficient is small.

\subsection{Fokker-Planck equation for AFM skyrmions}

Equation (\ref{sThiele2}) is stochastic, and it is difficult to solve it analytically. In this part, we find the steady-state velocity of AFM skyrmions by solving a deterministic Fokker-Planck equation related to the stochastic equation (\ref{sThiele2}).

A generic stochastic equation of motion can be written as,
\begin{align}
\label{Langevin}
\dot{m}_\alpha=g^{\alpha\beta}(\mathfrak{F}_\beta+\mathfrak{f}^{\mathrm{th}}_\beta),
\end{align}
where $g^{\alpha\beta}$ is the diffusion matrix; $\bm{\mathfrak{F}}$ and $\bm{\mathfrak{f}}^{\mathrm{th}}$ are the deterministic and stochastic forces, respectively; and the force autocorrelation function is $\langle \mathfrak{f}_\alpha^{\mathrm{th}}(\bm{r},t) \mathfrak{f}_\beta^{\mathrm{th}} (\bm{r}',t') \rangle=2\xi \delta_{\alpha\beta} \delta(\bm{r}-\bm{r}') \delta(t-t')$.
Let $P[\bm{m},t]$ be the probability of finding $\bm{m}$ at time $t$; then, the Fokker-Planck equation related to the above Langevin-like equation, Eq. (\ref{Langevin}), is given by \cite{FP},
\begin{align}
\label{FP}
\partial_t P=-\partial_\alpha(g^{\alpha\beta}\mathfrak{F}_\beta P)+\partial_\alpha\partial_\beta (\xi g^{\alpha\gamma}g^{\beta\gamma} P).
\end{align}

We can now find the Fokker-Planck equation related to the stochastic Thiele's equation (\ref{sThiele2}).
We consider a linear temperature gradient along the $x$-direction such that $\partial_y T=0$, $\partial^2_x T=0$, $J^m_y=0$ and $\partial_y \rho=0$; meanwhile, we assume that the magnon current density is almost uniform throughout the sample $\partial_x J^m_x=0$ and $\partial^2_x \rho=0$.
Thus, the components of reactive and dissipative forces, Eq. (\ref{forces}), as well as the diffusion matrix become,
\begin{align}
&F^\mathrm{r}_x=F^d_y=0, \\
&F^\mathrm{r}_y=-\frac{4\pi \hbar Q^n}{\Delta^2} {J}_x^n,\\
&F^\mathrm{d}_x=-\frac{c^2}{\Delta^2} M \partial_x \rho, \\
&g^{\alpha\beta}=-\frac{\gamma}{M \alpha a \mu_s}\delta_{\alpha\beta}.
\end{align}
The reactive force, ${\bm{F}}^\mathrm{r}$, has a component perpendicular to the AFM magnon current direction, while the dissipative force, ${\bm{F}}^\mathrm{d}$, is along the AFM magnon current. In AFM systems, the diffusion matrix, $g^{\alpha\beta}$, is diagonal and inversely proportional to the effective mass and damping parameter, while in FM systems, it has off-diagonal elements related to the {\it{magnetic}} topological number and diagonal elements proportional to the Gilbert damping \cite{kong2013dynamics,schutte2014inertia}.

The deterministic Fokker-Planck equation for AFM solitons becomes,
\begin{align}
&\partial_t P=-(g F^d_x-2g^2 \partial_x \tilde{\xi})\partial_x P-g F^r_y \partial_y P+g^2 \tilde{\xi}(\partial^2_x+\partial^2_y)P,
\end{align}
where $P(\bm{r},t)$ is the probability of finding the skyrmion at position $\bm{r}$ and time $t$. We are interested in the lowest-order traveling wave solution in the Fokker-Planck equation, thus defining $P=P(\bm{r}-\bm{v}t)$ and expanding to first order in the velocity; finally, we obtain,
\begin{align} \label{velocity}
v_x&=g F^d_x-2g^2 \partial_x \tilde{\xi}=\frac{\gamma c^2}{\alpha_\mathrm{G} a \Delta^2 \mu_s}\partial_x \rho-\frac{2\gamma^2 k_B}{M \alpha_\mathrm{G} a \Delta^2 \mu_s }\partial_x T\nonumber\\&\equiv v_x^n-v_x^B,\\
v_y&=g F^r_y=\frac{4\pi \hbar\gamma Q^n }{M \alpha_\mathrm{G} a\Delta^2 \mu_s} {J}_x^n\equiv v_y^n,
\end{align}
where $\bm{v}^n$ and $\bm{v}^B$ are the contributions from the AFM magnons and the stochastic Brownian motion, respectively. These two contributions have two opposite directions. In the low damping regime, the first term is dominant in large skyrmions and these large skyrmions move toward the hotter side. In small skyrmions, the second term is dominant and skyrmions move toward colder side of the system.
In AFM skyrmions, the dissipative torque is responsible for the longitudinal velocity, $v_x^n$, while in FM skyrmions, the longitudinal velocity arises from the adiabatic torque \cite{kong2013dynamics}. The transverse skyrmion velocity $v_y$ or skyrmion Hall velocity vanished in thermally driven skyrmion motion since thermal AFM magnons do not carry any net spin angular momentum ${J}_x^n=0$.

\subsection{Atomistic simulation}

\begin{figure}[t]\label{v-sky}
	\centering
	\includegraphics[width=0.5\textwidth]{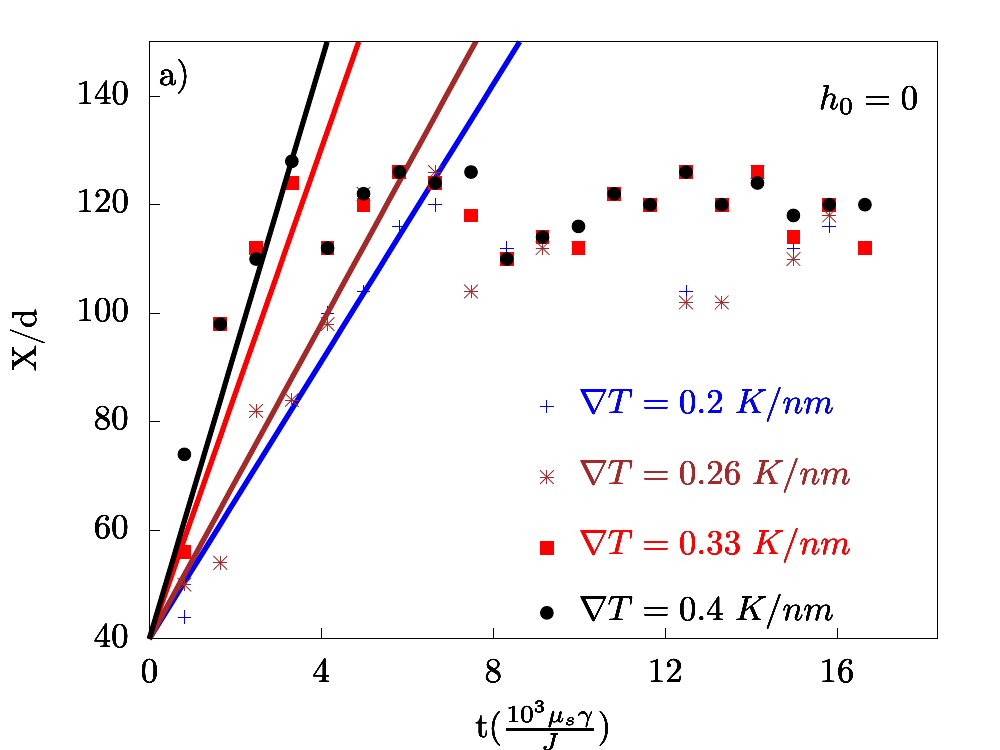}
	\hfill
	\includegraphics[width=0.5\textwidth]{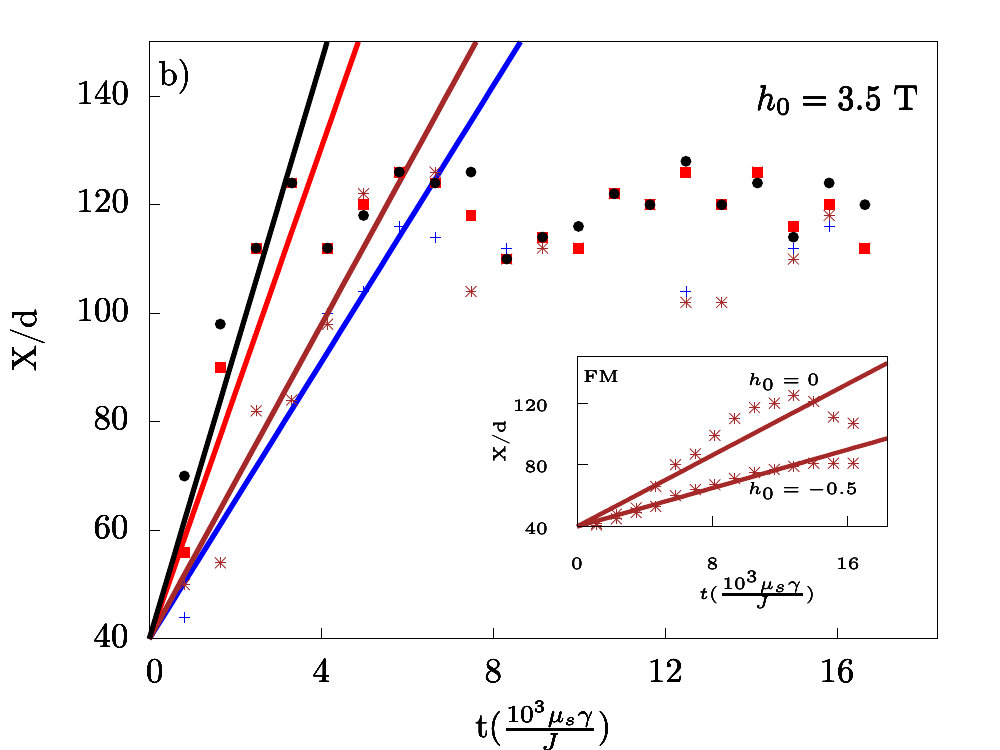}
	\caption{(Color online.) Skyrmion position as a function of time under different temperature gradients in the absence (a) and presence (b) of a uniform magnetic field. The inset shows the FM skyrmion velocity for both ${h_0}=0$ and ${h_0}=-0.5T$.}
\label{X-time}
\end{figure}

We simulate a 2D rectangular AFM system of $150d \times 50d$ with open boundary conditions and material parameters as $J=-5.44$ meV/atom, $D=0.18$ J, $K=0.1$ J, $\mu_s=2\mu_{B}$ and $\alpha_\mathrm{G}=0.07$. Within these material parameters a single skyrmion with a radius of $R/d \simeq 6$ can be created.
In the presence of the skyrmion at $(X_0,Y_0)=(40d,24d)$, a linear thermal gradient is applied along the $x$-direction, with $T(x=40d) < T(x=150d)$, and we trace the center of the skyrmion. Figs. \ref{X-time}-a and \ref{X-time}-b show the displacement of the skyrmion in the presence of different thermal gradients in the absence and presence of a perpendicular and uniform magnetic field, respectively. In the Supplemental Material \cite{suppl}, snapshots of the time evolution of skyrmion motion are presented.

The atomistic simulations show only a longitudinal displacement of AFM skyrmions in the presence of thermal magnons, as predicted by the analytical theory, $v_y^n=0$; see Eqs. (\ref{velocity}). Furthermore, also in good agreement with the theory, Eq. (\ref{velocity}), the skyrmion velocity is proportional to the temperature gradient.
Within the chosen parameters, the skyrmion is relatively large and moves toward the hotter region which means the velocity arising from the AFM magnon contribution is the dominant term  $v_x^n > v_x^B$. The effective interaction between the skyrmion and tilted spins at the boundary is repulsive \cite{iwasaki2013current} thus after some oscillations, the skyrmion lands at a distance from the rightmost edge (hotter side). Our atomistic simulations also show that the presence of external magnetic fields, less than the critical spin-flop field, has no significant effect on the AFM skyrmion velocity. This differs with respect to the dynamics of FM skyrmions, in which applying a magnetic field reduces the longitudinal skyrmion velocity; see the inset in Fig. \ref{X-time}-b.

By tuning the DMI and anisotropy, we can also create smaller skyrmions. Smaller AFM skyrmions are very unstable at finite temperatures. But those which have survived move toward the colder side of the system in the presence of an applied thermal gradient which means the Brownian contribution is dominant term $v_x^n < v_x^B$. In the Supplemental Material snapshots of the time evolution of skyrmion motion with a radius of $R/d \simeq 4$ are presented \cite{suppl}.

Here we should notice that in our simulations, we have assumed a very low Gilbert damping. Increasing the Gilbert damping leads to a drastic decay of thermal magnons through the system. In this case, there are many more magnons on one side of the skyrmion (the hotter side) than on the other side (the colder side). Consequently, this leads to a large
gradient of magnon number density and results in backward motion toward the hotter side even for smaller skyrmions, i.e., $v_x^n > v_x^B$.

\section{Summary and conclusion}

In summary, we have demonstrated a path for the ultrafast creation of single homochiral skyrmions via an effective magnetic field arising from the optical inverse Faraday effect. Since laser pulses are localized, the method facilitates the creation of skyrmions in a specific region, which makes it relevant to applications such as skyrmion-based synaptic devices \cite{Huang}.
The created single skyrmions are metastable states of a finite AFM system in the presence of DMI.

We have investigated the dynamic properties of AFM skyrmions via analytical calculations and classical atomistic simulations. The methods agree well. Thermal magnons move AFM skyrmions in a longitudinal direction; that is, the AFM skyrmion Hall angle is zero. In the low damping regime, large skyrmions move toward the hotter region, and small skyrmions move toward the colder side while in the large damping regime all skyrmins move toward the hotter side.  In addition, the AFM skyrmion velocity is much faster than for FM skyrmions under similar conditions.

{\it{Note added}} -- Recently, we became aware of another paper \cite{AFM-ani} that proposes a method for skyrmion motion in AFM insulators using a magnetic anisotropy gradient.

\section*{Acknowledgments}
We acknowledge fruitful discussions with J. Chico. The research leading to these results was supported by the European Research Council via Advanced Grant No. 669442, ``Insulatronics,'' and by the Research Council of Norway through its Centres of Excellence funding scheme, Project No. 262633, ``QuSpin.''

\end{document}